\documentclass[fleqn,10pt]{wlscirep}
\usepackage[utf8]{inputenc}
\usepackage[T1]{fontenc}

\usepackage{lineno}
\usepackage{booktabs}
\usepackage{multirow}
\usepackage{longtable}
\usepackage{xr-hyper}
\usepackage{xr}
\usepackage[charter,cal]{mathdesign}
\usepackage{enumitem}
\usepackage{xcolor}
\usepackage{soul}

\makeatletter
\newcommand*{\addFileDependency}[1]{
  \typeout{(#1)}
  \@addtofilelist{#1}
  \IfFileExists{#1}{}{\typeout{No file #1.}}
}
\makeatother

\newcommand*{\myexternaldocument}[1]{%
    \externaldocument{#1}%
    \addFileDependency{#1.tex}%
    \addFileDependency{#1.aux}%
}

\myexternaldocument{si}

\newcommand*\sfref[1]{%
    Supplementary Figure \ref{#1}}
\newcommand*\stref[1]{%
    Supplementary Table \ref{#1}}

\title{One country, multiple portraits: representativeness in GPS-based mobility data is source-specific and spatially dependent}

\author[a,b*]{Carmen Cabrera}
\author[a]{Francisco Rowe}
\author[a]{Miguel González-Leonardo}
\author[c]{Juan Ignacio Vilchis-García}
\author[d]{Elisa Omodei}
\author[e]{Maribel Hernández-Rosales}

\affil[a]{Geographic Data Science Lab, University of Liverpool}
\affil[b]{Centre for Advanced Spatial Analysis, University College London}
\affil[c]{El Colegio de Mexico}
\affil[d]{Department of Network and Data Science, Central European University}
\affil[e]{Center for Research and Advanced Studies of the National Polytechnic Institute, Irapuato Unit}

\affil[*]{Corresponding author: c.cabrera@liverpool.ac.uk}

\begin{abstract}

Anonymised GPS-based mobile phone data are increasingly used to estimate population distribution and human mobility, supporting applications across disaster response, public health, urban planning and migration research. Yet whether these data fairly represent the populations they describe, particularly outside high-income countries, remains poorly understood. We quantify coverage bias for 2,478 municipalities in Mexico by comparing population estimates from a single-platform source (Facebook) and a multi-app aggregator (Veraset) against the 2020 Mexican Population Census. We find that the magnitude and spatial distribution of coverage bias differ substantially across sources. Facebook provides higher and more evenly distributed coverage, whereas the multi-app data concentrate users in larger, wealthier and more digitally connected places. Coverage bias is also spatially structured, with neighbouring municipalities showing similar levels of over- or under-coverage. Using explainable machine learning, we show that digital access and material resources are the dominant drivers of bias for the multi-app data, while demographic and population structure dominate for Facebook. Explicitly modelling spatial dependence improves the performance of statistical models for explaining bias and reveals that an appreciable share of spatial variation remains unexplained by observed covariates. These findings show that coverage bias is source-specific and spatially dependent, and provide a foundation for adjustments that improve the representativeness of mobile phone data in unequal, data-scarce settings.
\end{abstract}

\begin{document}

\flushbottom
\maketitle


\section*{Significance Statement}
Censuses and household surveys, the traditional foundations of population data, are increasingly being supplemented or replaced by data passively collected from mobile phones, particularly in countries where official sources are infrequent or incomplete. But mobile phone data only see people who use particular devices and apps, raising the concern that these data systematically miss those least served by digital infrastructure. Studying coverage across 2,478 Mexican municipalities, we show that the answer depends sharply on which dataset is used, that errors cluster geographically in distinct regional patterns, and that these patterns are predictable from local conditions. Knowing where bias concentrates and what drives it is the foundation for correcting mobile phone data and using them more reliably.

\section*{Introduction}





Mobile phone data are increasingly used to estimate where people live and how they move, supporting applications across disaster response, public health, urban planning and migration research. Anonymised, GPS-based mobile phone data (MPD) record human presence and movement at spatial and temporal resolutions that traditional sources cannot match in near real time \cite{green2021}. Over the last decade, these data have become a valuable asset for tracking daily population rhythms \cite{gonzalez2008understanding}, commuting flows \cite{kung2014exploring, liu2020urban}, internal migration \cite{rowe2024medium, gubert2024temporary, blanchard2025highly} and population displacement after natural disasters \cite{rowe2022using} and conflict \cite{iradukunda2025estimating}. Their use is expanding fastest in lower- and middle-income countries \cite{rotondi2020leveraging, rowe2022methodological}, where censuses are infrequent, household surveys spatially coarse and administrative population registers often absent \cite{barreras2024exciting, rowe23-bigdata}. These are precisely where MPD can arguably make the largest difference in supporting policy by delivering more accurate population estimates.
 
Yet, a fundamental question about these data remains poorly answered. Do they fairly represent the populations they purport to describe? MPD only capture people who own a connected device, use a particular platform or application, and generate enough digital traces to be observed \cite{wesolowski13-biases, cabrera2025systematic}. Selection into the data tends to be shaped by income, age, urbanisation, digital infrastructure and platform-specific adoption \cite{suich2022coverage, wesolowski2016connecting}, and the resulting coverage biases can systematically distort inference if left unaddressed \cite{lai2019exploring, gallotti2024distorted}. Most empirical work using MPD treats these data as raw behavioural signals, with little attention to data bias and representativeness \cite{cabrera2025systematic}. Analyses that address bias in digital trace data more broadly typically rely on national-level calibrations that implicitly assume coverage error is homogeneous across a country and statistically independent across local areas \cite{ribeiro2020biased, gil2019demographic}. A smaller but growing literature has begun to validate MPD-derived population estimates against census benchmarks and finds substantial subnational variation in bias \cite{zhao2016understanding, milusheva2021assessing, li2024understanding, sanchez2026correcting, nijs2025data}. 

However, two limitations restrict what we can currently say. First, although bias in MPD is suspected to cluster geographically because its drivers do \cite{li2024understanding}, no existing diagnostic framework explicitly models this spatial dependence. This omission has two practical implications. When neighbouring areas carry information about each other's bias, treating them as independent observations discards a signal that could be used to detect bias in places where covariates on their own are uninformative, and to correct estimates where direct validation against ground truth is unavailable \cite{anselin1988spatial}. It can also inflate model performance when training and test sets contain spatially adjacent observations \cite{Roberts2017}. In such training scenarios, the model sees neighbours of the test cases during fitting, producing accuracy estimates that do not generalise to new unseen regions \cite{Legendre1989, Bahn2013, mahoney2023}. Both consequences impact on how reliably MPD can be used in places without recent census benchmarks. Second, the evidence base is drawn almost entirely from high-income, English-speaking countries \cite{nijs25, cabrera2025systematic, sinclair23}, leaving open whether bias-diagnostic frameworks built there transfer to settings with sharper inequalities in income and digital access. These settings are where MPD may be most useful but also most likely to misrepresent the population.
 
To address these gaps, we provide the first municipality-level assessment of coverage bias in GPS-based MPD for a Latin American country and use it to develop a spatially explicit framework for diagnosing bias more generally. We compare two data sources with fundamentally different generative processes: (i) a single-platform source (Facebook, via Meta's Data for Good initiative) reflecting selection into one mass-market application, and (ii) a multi-app aggregator (Veraset) reflecting participation in a broader ecosystem of location-enabled apps, against the 2020 Mexican Population Census across all 2,478 municipalities. Mexico is a particularly informative case. It combines pronounced regional inequalities in income, urbanisation and digital connectivity \cite{kim2016mexico, garcia2023exploring} with high-quality, fully enumerated census data, allowing us to test whether bias-diagnostic frameworks developed for high-income countries generalise to settings where MPD are most needed and most uneven. Methodologically, we extend the explainable machine-learning approach of \cite{cabrera2025systematic} by treating coverage bias as a spatial process. We test for spatial autocorrelation in bias, incorporate the spatial lag of bias into XGBoost models predicting it from local demographic, socioeconomic and digital-access characteristics, and evaluate models using spatial block cross-validation to avoid overstating generalisation. The result is a framework that quantifies how coverage bias varies across data sources, where it concentrates, what local conditions drive it, and how much of its spatial structure remains unexplained by observed covariates. Our framework provides an empirical and methodological foundation for the responsible use of digital trace data in unequal, data-scarce settings.


\section*{Results}

\begin{figure}
    \centering
    \includegraphics[width=0.95\linewidth]{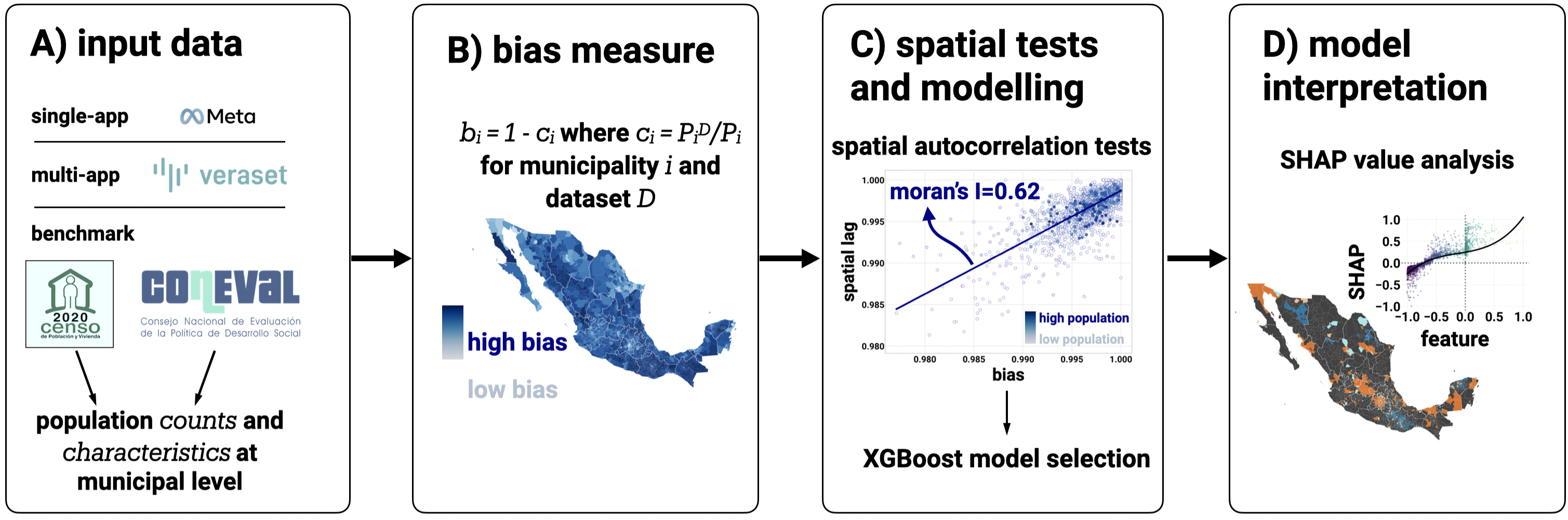}
    \caption{Overview of the analytical workflow used to assess bias in GPS-based mobile phone data (MPD). (A) Population estimates from single-app and multi-app MPD are analysed together with census-based variables to generate municipal-level population counts and covariates. (B) Bias $b_i$ is then estimated for each municipality $i$ and dataset $D$ as the deviation between MPD-based population $P_i^D$ and census population counts $P_i$. (C) Spatial autocorrelation in bias is assessed using global and local spatial diagnostics. Bias is modelled using XGBoost with spatial lag inputs and evaluated using spatial cross-validation. (D) Model outputs are interpreted using SHAP values to identify and compare the drivers and spatial patterns of coverage bias across GPS-based data sources.}
    \label{fig:overview}
\end{figure}

We analyse coverage bias in population estimates derived from two GPS-based mobile phone data sources and a census-based benchmark. Here, coverage bias refers to the extent to which each data source captures the underlying population across municipalities. Specifically, the analysis compares user-based population counts from a one-app platform (Facebook) and a multi-app mobile phone data source (provided by Veraset) against resident population counts from the 2020 Mexican Population Census, with bias assessed at the municipal level. Building on the framework introduced in \cite{cabrera2025systematic}, we quantify population coverage bias as a function the quotient between MPD-derived and census-based population counts, and examine how it varies across space and data sources. We then assess whether bias exhibits geographic clustering by using spatial autocorrelation tests to identify. Finally, we model coverage bias as a function of municipal-level socioeconomic, demographic and digital infrastructure characteristics, to identify the contextual drivers of over- and under-representation in MPD. We also test models that include the spatial lag of coverage bias as input, to examine whether information from neighbouring municipalities improves the explanation of bias beyond the observed covariates. Figure \ref{fig:overview} summarises the data inputs and analytical workflow. Further methodological details are provided in the Methods and Materials section.

\subsection*{The magnitude and distribution of coverage differ across datasets}

\begin{figure*}[!ht]
\centering
\includegraphics[width=0.8\textwidth]{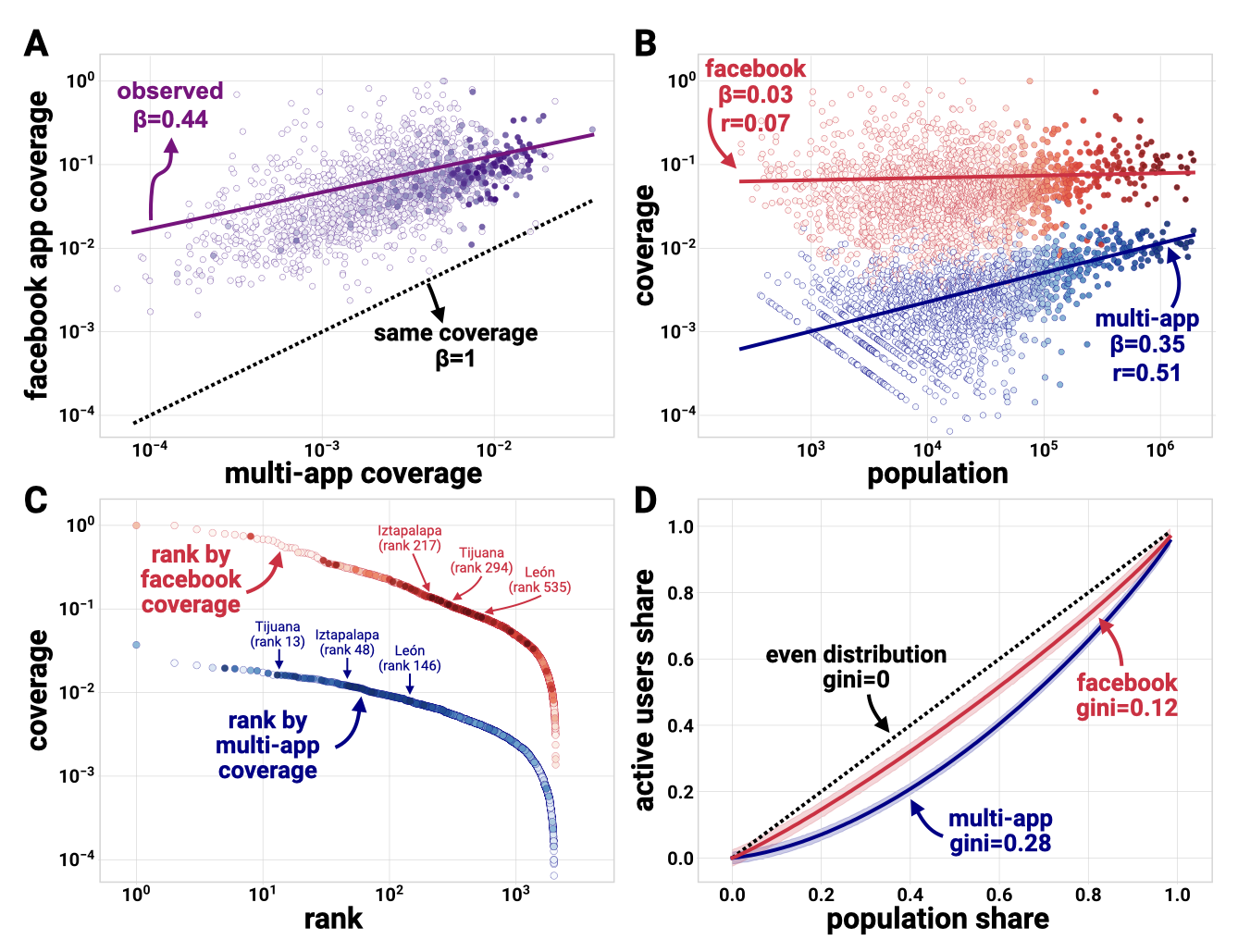}
\caption{Comparison of population coverage across single-app (Facebook) and multi-app GPS-based mobile phone datasets in Mexico. (A) Scatterplot comparing multi-app coverage (x-axis) and single-app coverage (y-axis) for each municipality, both on logarithmic scales; the dashed line denotes the $\beta = 1$ proportionality benchmark. (B) Relationship between municipal population size (x-axis) and coverage in each dataset (y-axis), both on logarithmic scales, with fitted scaling exponents $\beta$. (C) Rank distributions of municipalities ordered from highest to lowest coverage in each dataset, where darker colours indicate larger populations and the three highlighted points correspond to the most populous municipalities. (D) Lorenz curves showing the cumulative distribution of coverage relative to the cumulative share of population for each dataset, with Gini coefficients quantifying inequality.}
\label{fig:distribution}
\end{figure*}

We first compare population coverage in the single-app and multi-app GPS-based MPD sources across Mexican municipalities (Figure \ref{fig:distribution}). Throughout the analysis, population coverage is defined as the ratio between MPD-derived population estimates and resident population counts from the 2020 Mexican Population Census (see Methods and Materials section for further details). 

Data coverage diverges considerably in magnitude and distribution across data sources. This is illustrated in Figure \ref{fig:distribution}.A, where the x-axis shows the multi-app coverage for each municipality and the y-axis shows the corresponding Facebook (single-app) coverage. Each data point represents a municipality and the axes are in logarithmic scale. Darker colours represent larger populations. If both datasets provided equal population coverage across municipalities, observations would fall along the $y = \alpha x ^{\beta}$, with proportionality constant $\alpha=1$ and scaling exponent $\beta=1$ line (dashed line). Instead, the empirical relationship reveals that Facebook coverage increases slowly with the coverage provided by the multi-app data ($\beta = 0.44$). This indicates that municipalities that are well represented in the multi-app dataset do not receive proportionally higher representation in the Facebook dataset and \textit{vice versa}. 

Furthermore, the coverage observed in Facebook data is between one and two orders of magnitude higher than the coverage given by the multi-app source. This pattern likely reflects contrasting data-generating processes. Facebook is a mass-market platform with high penetration across demographic and socioeconomic groups in Mexico (83.5\% penetration rate as of January 2026, according to NapoleonCat \cite{napoleoncat26}), whereas the multi-app GPS data rely on opt-in location sharing through third-party applications, some of which may be for specific uses and audiences (e.g. assistance with translation, health and fitness tracking, weather prediction, etc.). In addition, the two datasets differ in how users are filtered, spatially aggregated and assigned a place of residence. Facebook data are smoothed and temporally aggregated as part of privacy-preserving procedures applied by Meta, while the multi-app data apply device-level filtering and residence inference based on sufficient nighttime GPS activity, which may exclude users with infrequent location pings (more details in the Methods section). 

We also find that coverage in the two datasets responds differently to population size, as the number of unique active users in multi-app data tends to be disproportionately higher as the underlying population size increases. This is illustrated in Figure \ref{fig:distribution}.B, where the x-axis shows municipal population and the y-axis shows the corresponding coverage from each dataset. Both axes are on a logarithmic scale and each data point represents a municipality, with darker colors capturing larger populations. If coverage was uniform regardless of the underlying population, we would expect a flat relationship, with Pearson's correlation coefficient $r$ and scaling exponent $\beta$ close to 0. The Facebook data (red) shows almost no association between coverage and population size ($\beta = 0.03$ and $r=0.07$), indicating that larger municipalities do not systematically receive proportionally greater coverage. By contrast, multi-app data displays non-linear scaling with population size ($\beta = 0.35$ and $r=0.51$). All the reported coefficients are significant with $p$-value $<0.05$. 

These patterns are further explored in Figure \ref{fig:distribution}.C and \ref{fig:distribution}.D, which demonstrate that Facebook data cover the population more evenly across municipalities, whereas multi-app data concentrate coverage within a smaller share of municipalities (i.e. more coverage for the population living in the most populated municipalities). Figure \ref{fig:distribution}.C shows the rank distribution of municipalities ordered from highest to lowest coverage, with darker colours indicating larger populations. This visualisation highlights how coverage is allocated across municipalities of different sizes. Figure \ref{fig:distribution}.D presents the corresponding Lorenz curves, which summarise the cumulative distribution of coverage relative to the cumulative share of population. These curves and the corresponding Gini coefficients are estimated from data following \cite{Sitthiyot21}. Greater deviations from the diagonal line reflect more uneven distributions and result in higher values of the Gini coefficient. We find that, for the multi-app dataset, the top-three most populated municipalities according to the census are highlighted (Iztapalapa, Tijuana and León), and they all have high ranks in the rank distribution shown in Figure \ref{fig:distribution}.C. This pattern reflects a more unequal allocation of coverage across municipalities according to their population size, consistent with the higher Gini coefficient of 0.28 observed in Figure \ref{fig:distribution}.D. By contrast, for the single-app dataset, high-population municipalities occupy relatively lower ranks in the rank distribution and coverage is more evenly distributed across municipalities of different sizes. This is reflected in a lower Gini coefficient of 0.12. These results suggest that population coverage displays some spatial dependence. We further explore these spatial patterns in the next subsection.

\subsection*{Spatial patterns in coverage bias also differ across datasets}

\begin{figure*}[!ht]
\centering
\includegraphics[width=0.8\textwidth]{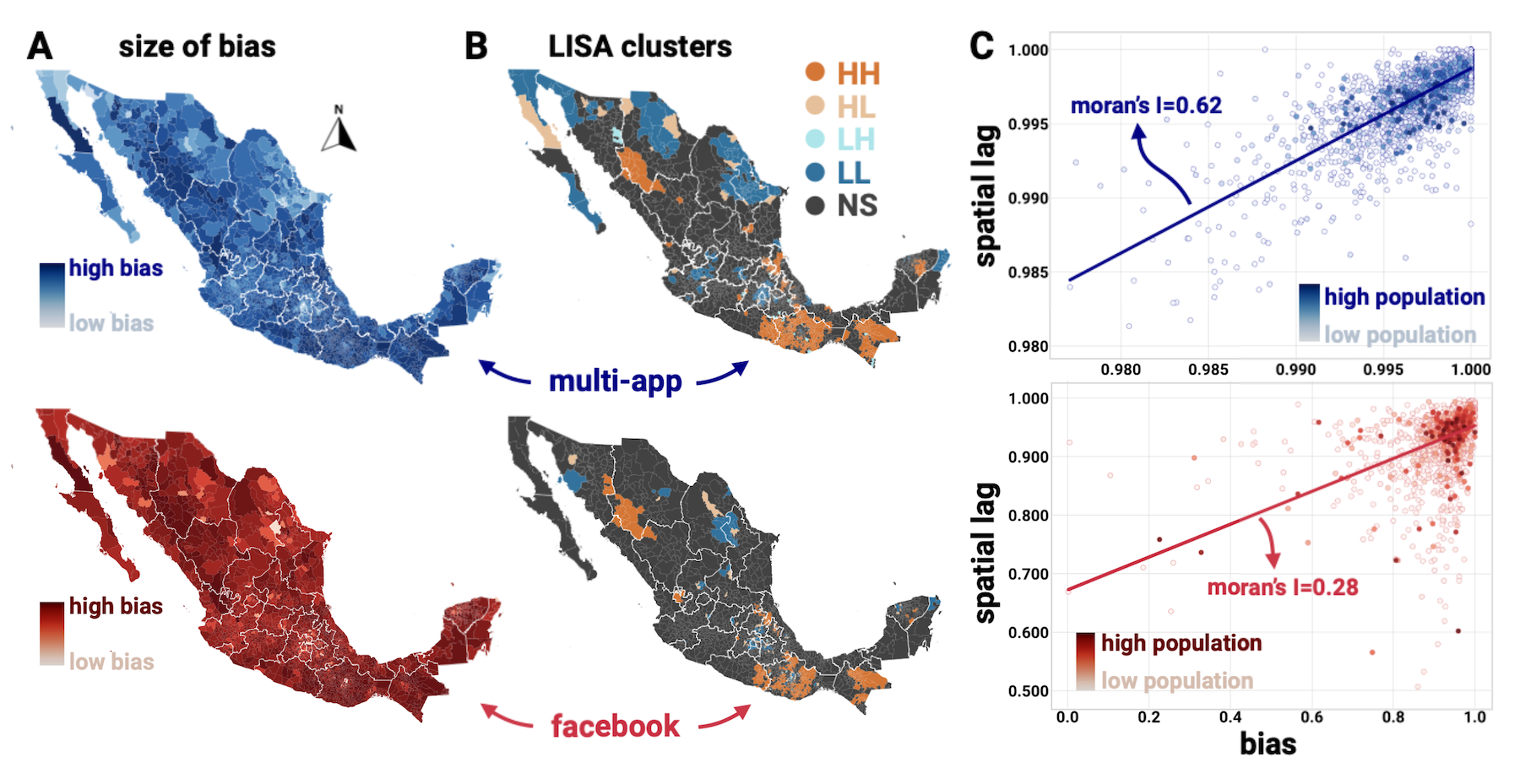}
\caption{Spatial patterns of coverage bias across municipalities for single-app and multi-app datasets. (A) Municipal-level maps of coverage bias ($b_i = 1-c_i$), with darker shades indicating higher bias (i.e. lower coverage relative to the census population). (B) Local Indicators of Spatial Association (LISA) cluster maps identifying high–high (HH), low–low (LL), high–low (HL), low–high (LH) and non-significant (NS) spatial clusters, with statistical significance assessed via permutation-based randomisation. (C) Moran plots showing global spatial autocorrelation in coverage bias, where each point represents a municipality and neighbouring municipalities are defined using a queen contiguity criterion \cite{Rey2023}. The slope of the fitted line corresponds to Moran’s $I$. Absolute values close to 1 indicate stronger spatial dependence.}
\label{fig:spatial}
\end{figure*}

We next examine whether coverage bias exhibits systematic spatial structure and whether these patterns differ across data sources. To do so, we examine a measure of coverage bias, which we define as $b_i = 1 - c_i$ for municipality $i$, where $c_i$ denotes the coverage ratio relative to the census population. Further details of the definition of this metric are provided in the Methods and Materials section. Figure \ref{fig:spatial} summarises the spatial distribution of bias across municipalities and demonstrates that coverage bias can form identifiable spatial clusters whose size and intensity vary considerably across datasets.

Coverage bias shows stronger spatial patterns in the multi-app dataset than in the single-app dataset. This is illustrated in Figure \ref{fig:spatial}.A-C. Figure \ref{fig:spatial}.A maps the magnitude of coverage bias across municipalities, with darker shades indicating higher bias (i.e. lower coverage relative to the census). These maps reveal that the multi-app dataset forms broad, contiguous regions of similar bias, whereas the spatial pattern in the Facebook dataset is more fragmented and dispersed. Figure \ref{fig:spatial}.B displays the corresponding Local Indicator of Spatial Association (LISA) cluster maps, which identify municipalities whose bias values are significantly similar (or dissimilar) to those of their neighbours based on permutation-based randomisation tests. The high–high and low–low clusters shown in these maps represent areas where bias is spatially concentrated, while high–low and low–high clusters indicate spatial outliers. The multi-app dataset generates larger, more coherent and numerous significant clusters, highlighting stronger spatial dependence, whereas the single-app dataset produces fewer and smaller clusters, reflecting weaker local spatial autocorrelation.

We numerically verify that the degree of spatial autocorrelation (i.e. the tendency for neighbouring municipalities to exhibit similar bias values) differs considerably across datasets and that it is higher for the multi-app dataset. These findings are shown in the Moran plots in Figure \ref{fig:spatial}.C. In these plots, bias for each municipality is plotted on the x-axis and the spatially lagged average bias of its neighbouring municipalities, defined using queen contiguity \cite{Rey2023}, is plotted on the y-axis. Points that align closely with the fitted regression line indicate municipalities whose bias values resemble those of their neighbours, contributing to positive spatial autocorrelation. The slope of this regression line corresponds to Moran’s $I$, a global index that quantifies the degree to which bias values cluster across space. For both datasets, the values of Moran’s $I$ are statistically significant at $p$-value $<0.05$, but their magnitudes differ. The multi-app dataset shows a steeper slope ($I=0.62$) and a tighter alignment along the regression line, indicating stronger spatial clustering of bias than for the single-app dataset ($I=0.28$).  

\subsection*{Evaluating bias modelling framework}

We model bias as a function of socioeconomic, demographic and digital accessibility covariates. These are described in the Methods and Materials section, and in \stref{tab:covariates}. We use an XGBoost model that extends the modelling framework of \cite{cabrera2025systematic}. Particularly, we compare XGBoost models estimated with and without the spatial lag of bias (the output variable) as one of the input features. This comparison allows us to assess whether information from neighbouring municipalities provides additional explanatory and predictive information beyond observed municipal covariates, indicating spatial structure not fully captured by the measured socioeconomic, demographic and digital-access variables. We evaluate model performance using block cross-validation \cite{mahoney2023}, which splits the area of analysis into a number of grid cells or ``blocks'' and then assigns all data into folds based on the blocks their centroid falls into. This choice of validation strategy is designed to avoid the over-optimistic validation results of purely random training–test splits in the presence of spatial autocorrelation, which can overstate generalisation to new areas \cite{Legendre1989, Bahn2013, Roberts2017, mahoney2023}. Further details of model specification, training and validation are given in the Methods and Materials section. 

\begin{figure*}[!ht]
\centering
\includegraphics[width=0.8\textwidth]{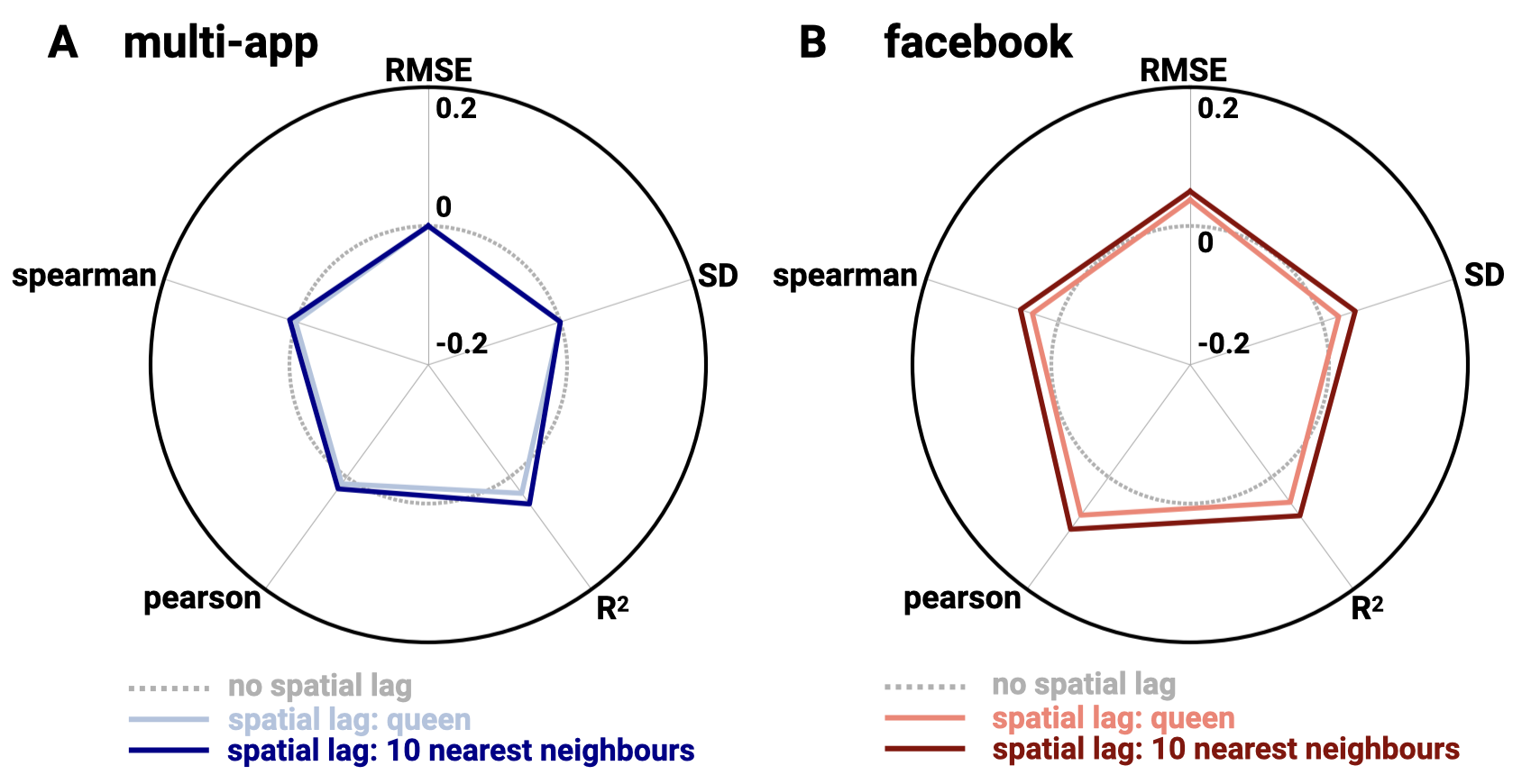}
\caption{Improved model performance after including spatially lagged bias as a model input suggests that neighbouring municipalities provide information about coverage bias beyond that captured by local demographic, socioeconomic and digital-infrastructure covariates. Radial plots summarise relative changes in model performance metrics for (A) the multi-app dataset and (B) the Facebook dataset when spatially lagged bias is included as an input feature in an XGBoost model, compared with a baseline specification without spatial lag. Metrics include root mean square error (RMSE), residual standard deviation (SD), $R^2$, and Pearson and Spearman correlations, shown relative to the baseline without spatial lag (dashed grey circle at 0). Values falling outside the dashed circle indicate improved performance (note that for RMSE, improvement corresponds to a reduction in error and is therefore plotted as a positive relative gain). Both spatial lag specifications, queen contiguity and 10 nearest neighbours, consistently improve model performance across metrics and datasets, with larger gains generally observed for the 10-nearest neighbours lag.}
\label{fig:performance}
\end{figure*}

We find that incorporating the spatial lag of bias as a model input consistently improves model performance across datasets. Figure~\ref{fig:performance} summarises relative improvements in performance metrics using radial plots for XGBoost models with spatial lag inputs compared with baseline models without spatial lags, for both the multi-app (Fig.~\ref{fig:performance}A)) and Facebook (Fig.~\ref{fig:performance}B) datasets. Improvements are observed across all evaluation metrics, including root mean square error (RMSE), residual standard deviation (SD), $R^2$, and Pearson and Spearman correlations, relative to baseline models without spatial inputs. For RMSE and standard deviation, improvements correspond to reductions in the value of the metric, but these are plotted as positive relative gains in performance. 

The gains in performance resulting from the inclusion of spatial autocorrelation as input indicate that coverage bias contains spatial signal that is not fully explained by the other model covariates. These performance gains are robust across alternative spatial lag specifications, including queen contiguity and 10-nearest neighbours, although the latter yields larger improvements, suggesting that spatial dependence in bias may extend beyond immediately adjacent municipalities. Robustness checks using random hold-out cross-validation with a 70–30 training–test split also yield consistent relative patterns. Full results are reported in \stref{tab:performance}. The magnitude of performance improvements is systematically larger for the Facebook models than for the multi-app models. This indicates that a greater share of spatial structure in Facebook coverage bias remains unexplained by the demographic, socioeconomic and digital access covariates. In contrast, in the multi-app data, coverage bias is more strongly correlated with these covariates, as shown in \sfref{fig:correlations}. As a result, the non-spatial models already account for a larger share of the spatial variation in bias, and less additional information is captured throught the includion of spatially lagged bias.

\subsection*{Comparative analysis of bias across GPS datasets}

\begin{figure*}[!ht]
\centering
\includegraphics[width=0.95\textwidth]{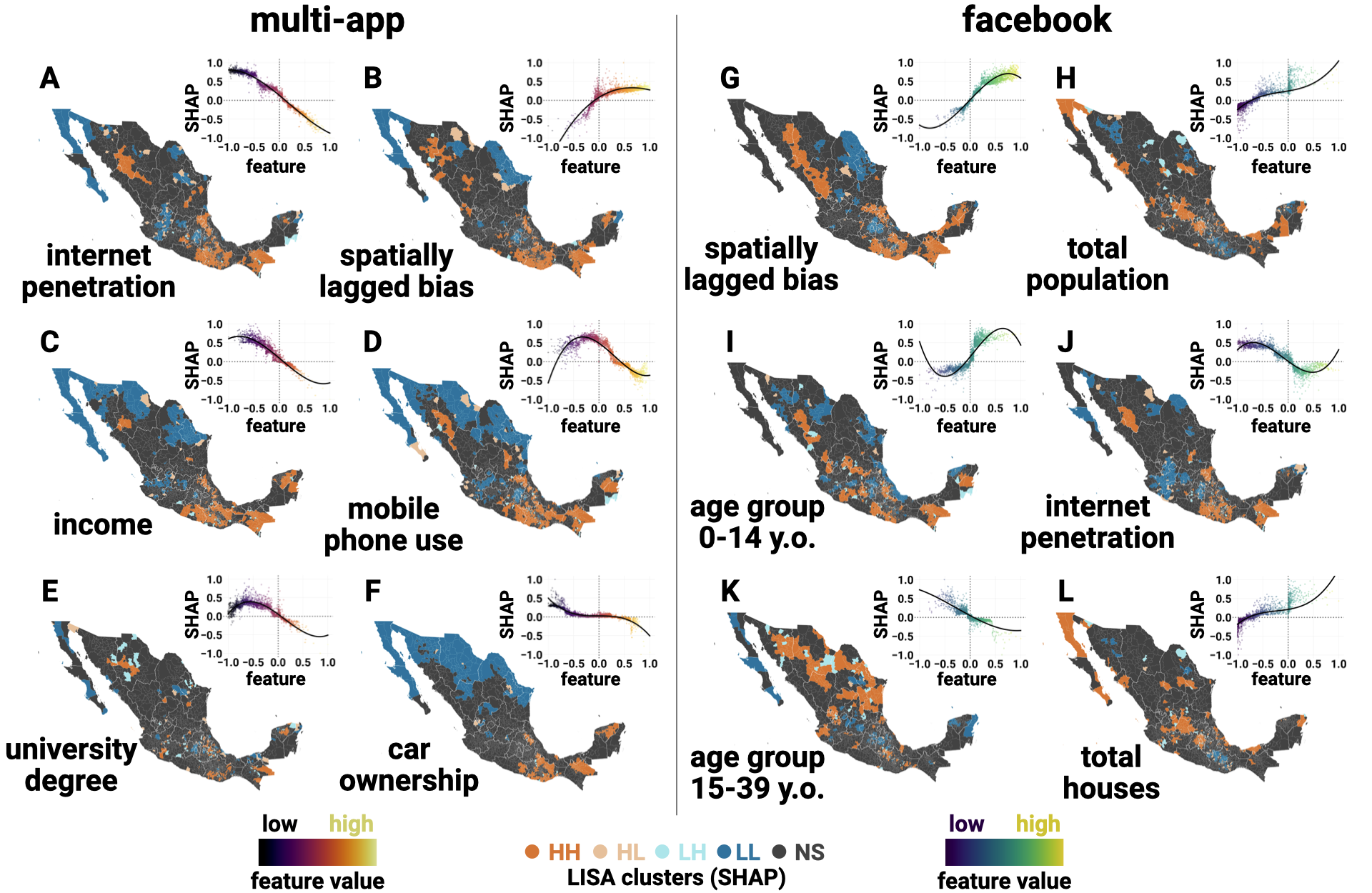}
\caption{Spatially explicit interpretation of XGBoost models using SHAP values. Maps show municipality-level SHAP values for the six most influential covariates identified by the XGBoost models for the multi-app (A–F) and single-app (G–L) datasets. Models are estimated using block cross-validation and include a spatially lagged bias term based on a 10-nearest neighbours specification. SHAP values represent the contribution of each covariate to the estimated coverage bias for a given municipality, relative to the average model prediction; positive SHAP values indicate that higher values of a given covariate are associated with increased estimated bias, whereas negative values indicate a bias-reducing contribution. LISA cluster labels identify spatial patterns in SHAP contributions, highlighting municipalities where similar covariate effects on coverage bias are spatially clustered (high–high or low–low) or dispersed (high–low or low–high). Non-significant spatial patterns are shown in grey. Insets depict SHAP dependence plots illustrating non-linear relationships between covariate values and their contributions to coverage bias.}
\label{fig5}
\end{figure*}

We compare the drivers of coverage bias across GPS-based datasets by interpreting the XGBoost models using SHAP values, which quantify the contribution of each covariate to the estimated bias at the municipal level. Figure~\ref{fig5} maps SHAP values for the six most influential covariates from models estimated using block cross-validation and including a spatially lagged bias term based on a 10-nearest neighbours specification. LISA diagnostics are used to identify municipalities where similar SHAP contributions are spatially clustered. Figure~\ref{fig5} also includes SHAP dependence plots, which illustrate the relationships between covariate values and their contributions to coverage bias, usually characterised by their non-linear behaviour.

We find that the relative importance of covariates differs across datasets, and these discrepancies are likely reflecting differences in the underlying data-generating processes. In the multi-app dataset (Fig.~\ref{fig5}A–F), coverage bias is most strongly associated with indicators of digital access and material resources, including Internet penetration, mobile phone use, income, education and car ownership. Higher values of these covariates are generally associated with reduced bias. This suggests that the data provide greater population coverage in municipalities with more access to digital technologies and higher socioeconomic status. The spatially lagged bias term also contributes to the model, but its relative importance is secondary (see Supplementary Fig.~\stref{fig:importance}, which reports covariate importance across model specifications). This pattern is consistent with earlier results showing that much of the spatial clustering in multi-app coverage bias is already aligned with socioeconomic and digital access characteristics.

In contrast, in the single-app dataset (Fig.~\ref{fig5}G–L), demographic structure plays a more prominent role, as variables such as total population size, total houses and age composition (particularly the shares of children and young adults) are identified as the most important drivers of bias. Internet penetration remains important, but its effect is weaker and more heterogeneous than in the multi-app data. This suggests that the adoption patterns of specific apps, such as Facebook, influence the relationship between Internet penetration and bias. The high importance of spatially lagged bias among the top predictors indicates that neighbouring municipalities provide key information for explaining Facebook coverage bias beyond the observed socioeconomic, demographic and digital-access variables. This suggests that the spatial patterns in bias reflect unmeasured regional processes or shared local contexts not captured by the rest of model covariates.

The relative importance of the top 20 covariates across model specifications with and without spatially lagged bias is shown in \sfref{fig:importance}, for both the multi-app and Facebook datasets and under both block cross-validation and random hold-out validation. This comparison shows that the inclusion of spatial lag does not explain away the influence of the demographic, socioeconomic and digital access covariates. In the multi-app models, indicators of digital access and socioeconomic status remain consistently among the most influential covariates across specifications. In the Facebook models, demographic and population-related covariates play a central role regardless of whether spatial lag is included, although the relative contribution of covariates becomes more evenly distributed when spatial dependence is explicitly modelled. The overall ranking and composition of influential covariates are stable across validation strategies. 

\section*{Discussion}

Coverage bias in GPS-based mobile phone data is source-specific and spatially structured, varying systematically across data sources and across geographic locations. Treating coverage bias as uniform across a country or as independent across local areas, as commonly done in current applications, produces a misleading picture of population representativeness and limits the reliability of mobile phone data for research and policy.

Three lines of evidence support this conclusion. First, we show that the two GPS-based sources we examined (a single-platform feed from Facebook and a multi-app aggregator) produce markedly different portraits of the same country. Facebook covers Mexico's municipalities at consistently higher rates and more evenly across population sizes, while the multi-app data concentrate users in large, wealthier and more digitally connected places. These differences are not noise; they reflect distinct selection mechanisms, with Facebook reaching a mass-market user base and the multi-app feed depending on opt-in location sharing across smaller, more specialised applications. Second, coverage bias is not randomly distributed across space. in both sources, coverage bias forms statistically significant geographic clusters, but the strength of clustering differs: Moran's $I$ is more than twice as large in the multi-app data ($I=0.62$) as in Facebook ($I=0.28$), and LISA diagnostics reveal that the two sources cluster bias in different parts of the country. Third, modelling bias as a function of local demographic, socioeconomic and digital-access characteristics reproduces some (but only some) of this spatial structure. Adding the spatial lag of bias as a predictor improves model performance across every metric we examined, and does so most strongly for the Facebook data, indicating that an appreciable share of spatial variation in bias reflects regional processes not captured by observed covariates. 

The drivers of bias also differ across sources in ways consistent with their generative processes. In the multi-app data, the strongest predictors are indicators of digital access and material resources (particularly internet penetration, mobile phone use, income, education and car ownership) and higher values of these covariates reduce bias. In the Facebook data, demographic structure (population size, total houses, age composition) carries more weight, with internet penetration playing a secondary and more heterogeneous role. This asymmetry mirrors a deeper feature of the Mexican context. Mexico is a country where GDP per capita varies by more than an order of magnitude across regions \cite{busso20} and where access to smartphones and reliable internet is strongly stratified by income, education and urbanisation \cite{martinez2020internet, diaz2021factors, Garcia-Mora25}. This digital divide is reproduced in the data. Notably, this differs from the United Kingdom, where digital access is more strongly patterned by demographic than socioeconomic factors \cite{cabrera2025systematic}, suggesting that bias-correction strategies developed in one context cannot easily be transferred to another if structural societal differences exist.

Our work has important implications for the use of GPS-based mobile phone data in official statistics and policy-relevant research. Common practice is to validate that aggregate user counts are roughly proportional to known population counts at coarse spatial scales \cite{renninger26, wang22, moro2021, cabrera25latin}. Our findings revealed this is not sufficient. Two datasets can pass such a check while differing systematically in which places and which people they over- or under-represent, and the structure of those errors is spatial. We therefore argue that coverage diagnostics should be source-specific and spatially explicit. MPD biases should be evaluated more systematically before these data are used in substantive applications, disclosing the extent to which population groups and places are represented in the data. Such evaluation should be reported with the findings as part of a transparent assessment of data quality.

Despite the advances made here in the study of coverage bias in mobile phone data, outstanding questions remain. First, the extent and spatial patterns of bias are likely to vary across countries depending on levels of socioeconomic inequality, digital access, platform penetration, urbanisation and governance \cite{ribeiro20-facebook, gil-clavel2019, casey25, demography25mexico}, and comparative evidence to date is limited and largely focused on gender \cite{casey25}. Second, coverage bias also has a temporal dimension that remains under-studied, with existing evidence confined to particular platforms or short time windows \cite{sanchez2026correcting, li24biasUS, duan24counterurb, cabrera25latin}. Third, although progress has been made in measuring and assessing bias in aggregated MPD \cite{cabrera2025systematic, nijs25}, methods to adjust mobile phone data for inference of population counts \cite{andrich2026} and mobility flows are still needed, and need to be generalisable across digital sources and reliable enough for use in official statistics and policy applications.

Advancing this research agenda on the assessment and adjustment of bias in MPD requires high-quality benchmark datasets against which coverage bias can be evaluated. Yet such benchmark data are unevenly available across countries. For example, in some contexts, censuses are outdated, inaccessible, insufficiently spatially granular or subject to concerns about quality and coverage \cite{pelletier2020census, Kukutai25}. Resources such as WorldPop \cite{worldpop-dataset}, the European Commission's Global Human Settlement Layer \cite{ghsl-dataset} and similar data products provide a useful basis for validation by offering spatially granular and regularly updated estimates of population distribution, particularly for locations where official benchmark data are limited. We therefore call for new initiatives to develop benchmark resources that combine fine-grained estimates for population counts and flows with richer sociodemographic covariates and spatially explicit indicators of digital technology access and use. Such investments are essential to move the field beyond acknowledging data limitations and towards operational solutions, reproducible correction frameworks, validation protocols, uncertainty reporting and stronger norms of transparency.

\section*{Methods and Materials}

\subsection*{Data sources and pre-processing}

\subsubsection*{GPS-based Meta-Facebook data} 

We used aggregated and anonymised GPS-based mobile phone data from Meta’s Data for Good Initiative. This dataset contains geolocated data from Facebook smartphone app users with the location sharing option turned on. To ensure privacy, Meta removes users’ individual characteristics, aggregates the data and applies three privacy-preserving techniques. Particularly, counts involving fewer than 10 users are removed, a small amount of random noise is added to the count data and spatial smoothing is applied \cite{maas19}. While these techniques protect user privacy, they may lead to underrepresentation in sparsely populated areas \cite{cabrera25latin}.

The dataset contains the number of active app users for each cell of the Bing Maps Tile System, developed by Microsoft \cite{microsoftBingMaps}. This system includes partitions of the world map in the form of square cells with varying levels of resolution. Our dataset uses Level 13 Tiles for Mexico, consisting of cells of approximately 6 square kilometres, although the size varies slightly according to distance from the Equator. The dataset provides information for three daily 8-hour time windows (i.e. 00:00-08:00, 08:00-16:00, and 16:00-00:00) in Pacific Time. Each user's location is defined as the tile in which they spent most of their time during a given time window. To estimate the users’ place of residence, we restricted our analysis to the night-time interval (00:00-08:00, or 02:00-10:00 local time), when people are most likely to be at home, in line with other home-detection algorithms \cite{wang22, Verma2024, Zhong2025}. For spatial and temporal comparison with the Mexican Census, we computed a daily average user count and then aggregated the resulting averages to the municipal level. As a robustness check, we also reverse the order of aggregation by first spatially aggregating user counts to municipalities for each day and subsequently computing temporal averages. The resulting number of counts by municipality remains practically the same, as demonstrated in \sfref{fig:tts-stt}.

The dataset was developed by Meta to support the analysis of population movements during the COVID-19 pandemic. It includes information for the pandemic period (mid-April 2020 onward), as well as a pre-pandemic baseline corresponding to the 45 days prior to 10 March 2020. The baseline was computed by averaging observations for each day of the week and each time window over the entire 45-day period. Baseline values are included in datasets released from mid-April 2020 onward and correspond to the same day of the week and time window as those observed during the pandemic period. To obtain a consistent daily estimate of Facebook users corresponding to the closest period available to the benchmark date of the 2020 Mexican census (i.e. 15 March 2020), we used baseline values corresponding to 106 million observations, from datasets released between 2 April 2020 and 12 April 2020.

\subsubsection*{Multi-app GPS-based mobile phone data} 

We used multi-app GPS-based mobile phone data for Mexico provided by the company Veraset. The company’s raw mobility data originate from anonymised geolocation signals collected from mobile devices through partnerships with third-party mobile applications, software development kits and data aggregators that obtain user consent for location access. These signals consist of timestamped latitude-longitude pings generated when a device interacts with GPS. Once acquired, Veraset aggregates and filters the data to remove noise and low-quality points, standardises the records and assigns each ping to a unique but anonymised device identifier. The resulting dataset represents high-frequency, device-level mobility traces that can be used to infer movement patterns, visits to points of interest and broader population-level mobility dynamics, while maintaining the anonymity of individual users.

To estimate the place of residence of mobile devices, we processes the geolocation records to obtain, for each unique device, the municipality with the highest number of records during nighttime hours (i.e. 00:00-05:00). The resulting municipality is assigned as the place of residence. Then, the total number of devices per municipality for each day of the study period are counted, totalling approximately 24 million device-week observations. To obtain a consistent daily estimate that aligns closely with the benchmark date of the 2020 Mexican census, we used data for a full week (from the 15th March 2020 to the 22nd March 2020) and calculated a daily average.
   
\subsubsection*{Census-based population data}

We used data from the 2020 Mexican Census, which covers the entire national population, to compared total population counts from the census with those from Facebook and multi-app mobile phone data across Mexico’s 2,478 municipalities. This approach enables us to calculate the spatial coverage of both digital sources (i.e., the number of users divided by the census population in each municipality).

\subsubsection*{Covariates from Census and CONEVAL data}

We selected a set of sociodemographic variables from the 2020 Mexican Census (see INEGI, 2021) as predictors to investigate the underlying factors explaining spatial biases across municipalities in the coverage of Facebook and multi-app mobile phone data. The full list of predictor variables is reported in \stref{tab:covariates}. Specifically, we selected the following variables, which we grouped into three blocks:
\begin{itemize}[noitemsep, topsep=0pt]
    \item Access to technology: percentage of mobile phone users and households with internet access.
    \item Demographic variables: number of households, total population, mean age, percentage of women, percentage of individuals aged 0-14, 15-39 and 65+, percentage of immigrants, percentage of afro-descendants, percentage of indigenous-language speakers, and percentage of population with disabilities.
    \item Socioeconomic variables: income, mean number of occupants per household, percentage of households with electricity, literacy rate, percentage of working population, percentage of individuals with a university degree, and percentage of households with a car.
\end{itemize}

We also included the following socioeconomic variables from the Mexican National Council for the Evaluation of Social Development Policy (CONEVAL): percentage of the population living in poverty, percentage living in extreme poverty, percentage with income vulnerability, percentage of housing units in poor condition, and percentage of the population experiencing educational lag. See CONEVAL (2021) for details on the construction of these variables.

\subsection*{Measuring population coverage bias}

We define a metric to quantify the magnitude of population coverage bias in each Mexican municipality. The metric is based on the population coverage of a given MPD source, computed as the ratio between the population estimated from MPD and the resident population reported by the census for the same geographic unit. Let $P_i^D$ denote the MPD-derived population estimate for municipality $i$ from data source $D$, and let $P_i$ denote the corresponding resident population count from the 2020 Mexican Population Census. Population coverage in municipality $i$, denoted $c_i$, is defined as:
\begin{equation}
c_i = \frac{P_i^D}{P_i}.
\end{equation}

The coverage ratio $c_i$ represents the proportion of the resident population captured by the MPD source. Values of $c_i = 1$ indicate full population coverage, whereas values below 1 indicate under-coverage. In some cases, coverage ratios may exceed 1 if the MPD-derived population estimate exceeds the census population, for example due to multiple device ownership or multiple application accounts per individual.

We define population coverage bias in municipality $i$ as the deviation from full coverage:
\begin{equation}\label{eq:bias}
b_i = 1 - c_i.
\end{equation}

A value of $b_i = 0$ indicates no coverage bias, corresponding to full population coverage, while larger absolute values indicate increasing levels of underrepresentation. We use this bias metric to analyse the magnitude, spatial distribution and variability of coverage bias across Mexican municipalities.

\subsection*{Spatial analysis of coverage bias}

To examine whether population coverage bias exhibits spatial dependence and whether this dependence differs across data sources, we conduct a spatial autocorrelation analysis of coverage bias at the municipal level. Coverage bias for municipality $i$ is defined as $b_i = 1 - c_i$, where $c_i$ denotes the population coverage ratio relative to the 2020 Mexican Population Census (see equation \eqref{eq:bias}). 

Spatial dependence is assessed using both global and local measures of spatial autocorrelation. Spatial relationships between municipalities are represented using a contiguity-based spatial weights matrix $\mathbf{W}$ constructed according to a queen contiguity criterion, whereby two municipalities are considered neighbours if they share either a boundary or a vertex. The weights matrix is row-standardised so that the weights associated with each municipality sum to one. To assess the robustness of the results to the choice of spatial weights, we also estimate spatial autocorrelation statistics using alternative neighbourhood definitions, including rook contiguity and distance-based weights. The corresponding results are reported in \stref{tab:spatial-stats}.

Global spatial autocorrelation in coverage bias is quantified using Moran’s $I$, which measures the extent to which municipalities with similar bias values are spatially clustered. Moran’s $I$ is defined as:
\begin{equation}
I = \frac{n}{\sum_{i}\sum_{j} w_{ij}} 
\frac{\sum_{i}\sum_{j} w_{ij}(b_i - \bar{b})(b_j - \bar{b})}
{\sum_{i}(b_i - \bar{b})^2},
\end{equation}
where $n$ is the number of municipalities, $b_i$ is the coverage bias in municipality $i$, $\bar{b}$ is the mean coverage bias across all municipalities, and $w_{ij}$ denotes the $(i,j)$ element of the spatial weights matrix $\mathbf{W}$. Statistical significance is assessed using permutation-based randomisation tests with 999 permutations, in which bias values are randomly reassigned across municipalities to generate a reference distribution under the null hypothesis of spatial randomness.

To identify localised spatial clustering and spatial outliers, we compute Local Indicators of Spatial Association (LISA). The local Moran statistic for municipality $i$ is given by:
\begin{equation}
I_i = (b_i - \bar{b}) \sum_{j} w_{ij}(b_j - \bar{b}),
\end{equation}
which measures the similarity between the bias in municipality $i$ and the average bias of its neighbouring municipalities. Based on the sign and statistical significance of $I_i$, municipalities are classified into high-high, low-low, high-low or low-high clusters. Statistical significance is again evaluated using permutation-based tests at the 5\% level.

\subsection*{Modelling population coverage bias}

To identify the contextual factors associated with population coverage bias and to assess the importance of accounting for spatial dependence, we model municipal-level coverage bias as a function of demographic, socioeconomic and digital access characteristics using explainable machine learning. Existing evidence shows that population data derived from digital platforms tend to overrepresent urban, wealthier and younger populations, reflecting uneven access to and engagement with digital technologies \cite{blumenstock2010, wesolowski13-biases, schlosser21-biases}. We conceptualise population coverage bias as arising from systematic differences in local population characteristics and technological access rather than as random measurement error. Our outcome variable is the coverage bias indicator $b_i$, defined as the deviation between MPD-derived population estimates and census-based resident population counts for municipality $i$, and the covariates are measured using data from the 2020 Mexican Population Census and complementary administrative and geospatial sources. The full list of predictors is reported in \stref{tab:covariates}.

To model the relationship between covariates and the outcome, we build on the approach introduced in \cite{cabrera2025systematic} by using an eXtreme Gradient Boosting (XGBoost) regression framework. We extend the original approach by adopting spatial block cross-validation during model training and evaluation and by comparing baseline models with specifications that additionally include spatially derived covariates. XGBoost is an ensemble learning method that combines predictions from multiple regression trees to produce a flexible, nonlinear model. Trees are constructed sequentially, with each successive tree fitted to minimise the residual errors of the existing ensemble. The XGBoost regression model is specified as:
\begin{equation}
\widehat{b}_i 
= \sum_{m=1}^{M} f_m\bigl(X_i\bigr),
\quad f_m \in \mathcal{F},
\end{equation}
where $b_i$ denotes the observed population coverage bias in municipality $i$, $\widehat{b}_i$ is the predicted bias, $f_m$ represents an individual regression tree from the ensemble $\mathcal{F}$, $M$ is the total number of trees and $X_i$ denotes the vector of municipal-level covariates. The model iteratively learns the contribution of each covariate to coverage bias, allowing for complex nonlinear effects and interactions across predictors.

As a baseline specification, we estimate an explainable XGBoost model using the covariates listed in \stref{tab:covariates} as inputs. Models are trained and evaluated using spatial block cross-validation \cite{Valavi19, mahoney2023}, which partitions municipalities into spatially contiguous folds. This validation strategy mitigates over-optimistic performance estimates that can arise under random training–test splits in the presence of spatial autocorrelation. Model hyperparameters, including learning rate, maximum tree depth, subsampling ratios and regularisation penalties, are tuned using grid search and cross-validation on the training data. We apply both L1 (Lasso) and L2 (Ridge) regularisation penalties to control model complexity, promote covariate sparsity and improve generalisation. The tree-based structure of XGBoost further reduces sensitivity to multicollinearity by hierarchically selecting informative splits. Final models are fitted using the hyperparameter configuration that minimises the root mean squared error (RMSE). This hyperparameter configuration is reported in \stref{tab:hyperparameters}.

To assess whether neighbouring areas provide additional information for explaining coverage bias beyond observed covariates, we also estimate ``spatially aware'' XGBoost models for each dataset by augmenting the baseline feature set with the spatial lag of the outcome variable. The spatial lag is defined as the weighted average of coverage bias observed in neighbouring municipalities and is constructed using the same spatial weights matrix $\mathbf{W}$ described in the spatial autocorrelation analysis. We consider two alternative neighbourhood definitions when constructing $\mathbf{W}$: (i) a contiguity-based Queen neighbourhood scheme and (ii) a distance-based $k$-nearest neighbours (kNN) scheme.

Model performance is evaluated using complementary metrics, which include root mean squared error (RMSE), Pearson and Spearman correlation coefficients, the standard deviation of prediction errors, and the coefficient of determination ($R^2$). To check the robustness under the choice of model validation strategy, we also estimate baseline and ``spatially-aware'' models using random holdout validation \cite{}, in which 70\% of municipalities are used for model training and the remaining 30\% are reserved for out-of-sample testing. Results from the random holdout validation are reported in \stref{tab:performance}, which demonstrates that model performance is consistent across validation strategies.

\subsubsection*{Model comparison and interpretation}

To interpret model outputs and identify the most influential predictors of coverage bias, we employ explainability tools based on SHapley Additive exPlanations (SHAP). SHAP values provide consistent measures of the marginal contribution of each predictor to the model predictions, allowing to quantify the direction and magnitude of the contribution of each covariate to estimated coverage bias across municipalities. We summarise global feature importance by aggregating absolute SHAP values across observations, enabling a comparison of the relative influence of demographic, socioeconomic and digital access covariates across model specifications. 

\sfref{fig:importance} reports the rank and relative importance of the top 20 covariates for each dataset, comparing models estimated with and without spatially lagged bias and under both block cross-validation and random hold-out validation. In the main text, we focus on the six most influential covariates, which are presented in Figure~\ref{fig5} for models estimated using block cross-validation and including spatial lag. In addition, we use SHAP dependence plots to examine how estimated coverage bias varies with changes in individual covariates, while accounting for nonlinear effects and interaction patterns learned by the model. These analyses allow us to assess whether incorporating spatial dependence alters both the ranking of influential covariates (\stref{fig:importance}) and the functional form of their relationships with coverage bias (Fig.~\ref{fig5}), thereby informing substantive conclusions about the contextual drivers of coverage bias in GPS-based mobile phone data.

\bibliography{references}

\section*{Acknowledgements}

CC and FR acknowledge the financial support of the UK's Economic and Social Research Council (project ID: ES/Y010787/1).

\section*{Author contributions statement}

\section*{Data and code availability}

The code is openly available at \href{https://github.com/Vilchis-17/DEBIAS_MX}{https://github.com/Vilchis-17/DEBIAS\_MX}.

\section*{Competing interests}
The authors declare no conflict of interest.

\end{document}



\flushbottom
\maketitle

\begin{table}[h!]
    \centering
    \includegraphics[width=0.9\linewidth]{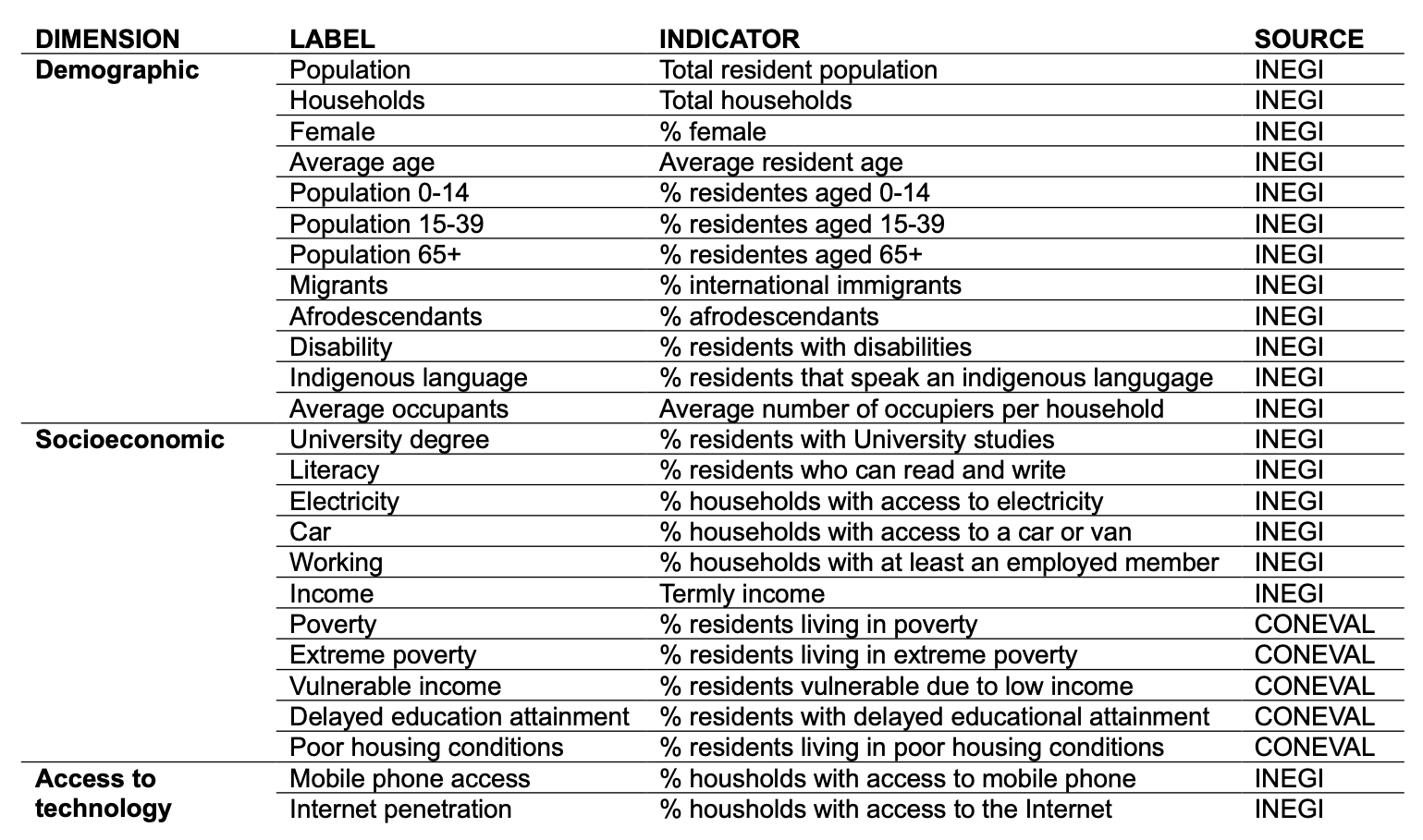}
    \caption{Municipal-level covariates used for modelling coverage bias, classified into demographic, socioeconomic and digital technology access. Demographic and technology access indicators are derived from the 2020 Mexican Population Census produced by the National Institute of Statistics and Geography (INEGI). Poverty and deprivation measures are obtained from the National Council for the Evaluation of Social Development Policy (CONEVAL).}
    \label{tab:covariates}
\end{table}


\begin{table}[h!]
    \centering
    \includegraphics[width=0.9\linewidth]{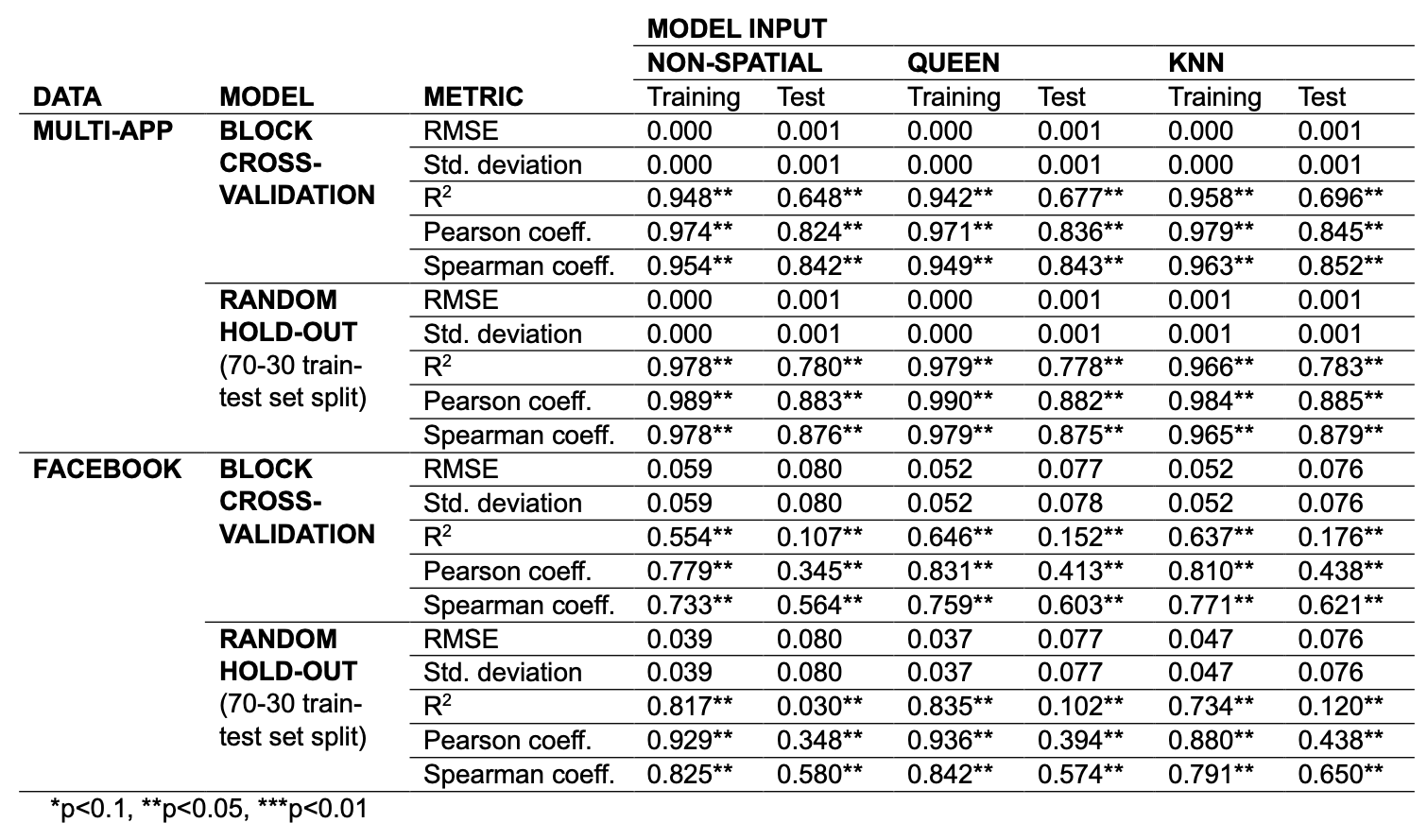}
    \caption{Performance metrics for coverage bias models under alternative validation strategies and spatial specifications, reported separately for the multi-app and Facebook datasets. Results are shown for block cross-validation and random hold-out validation with a 70–30 training–test split. For each validation strategy, model performance is reported for a non-spatial specification and for other specifications that include a spatial lag of the outcome based on queen contiguity and k-nearest neighbours (kNN). Metrics include root mean squared error (RMSE), residual standard deviation, coefficient of determination ($R^2$), and Pearson and Spearman correlation coefficients, reported separately for training and test samples. Statistical significance levels are indicated by asterisks.}
    \label{tab:performance}
\end{table}

\begin{figure}[h!]
    \centering
    \includegraphics[width=0.7\linewidth]{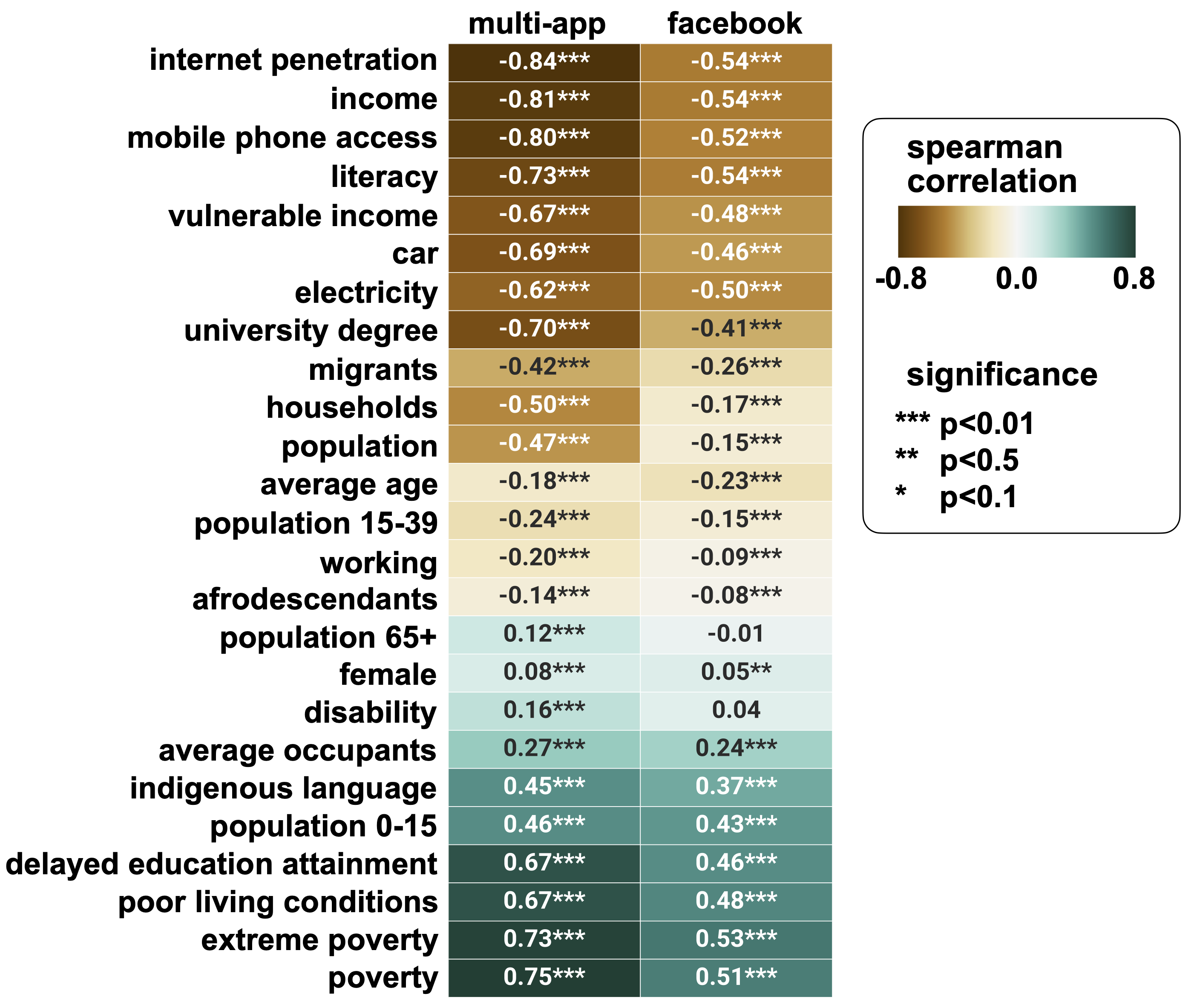}
    \caption{Pairwise correlations between municipal-level coverage bias and demographic, socioeconomic and technology access model covariates, for the multi-app and Facebook datasets. Correlations are measured with Spearman's correlation coefficient. Colour intensity indicates the magnitude and direction of the correlation, with negative values corresponding to lower bias (higher coverage) in municipalities with higher covariate values. Statistical significance is assessed using two-sided tests and indicated by asterisks ($^* p<0.1$, $^{**} p<0.05$, $^{***} p<0.01$). All variables are measured at the municipal level.}
    \label{fig:correlations}
\end{figure}

\begin{figure}[h!]
    \centering
    \includegraphics[width=0.7\linewidth]{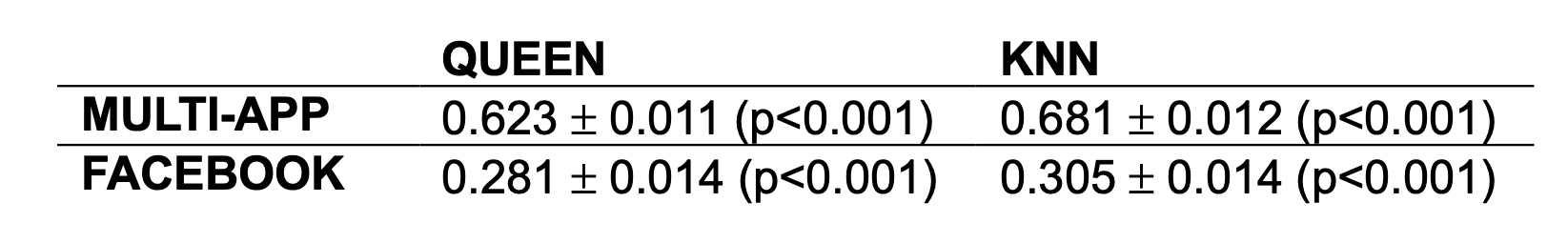}
    \caption{Global Moran’s $I$ statistics for municipal-level coverage bias, reported for the multi-app and Facebook datasets using queen contiguity and $k$-nearest neighbours (kNN) spatial weight specifications. Reported values correspond to the mean Moran’s $I$ across permutations, with standard deviations in parentheses. Statistical significance is assessed using permutation-based tests.}
    \label{tab:spatial-stats}
\end{figure}

\begin{landscape}
\begin{figure}
    \hspace*{-0.05\linewidth}
    \includegraphics[width=\linewidth]{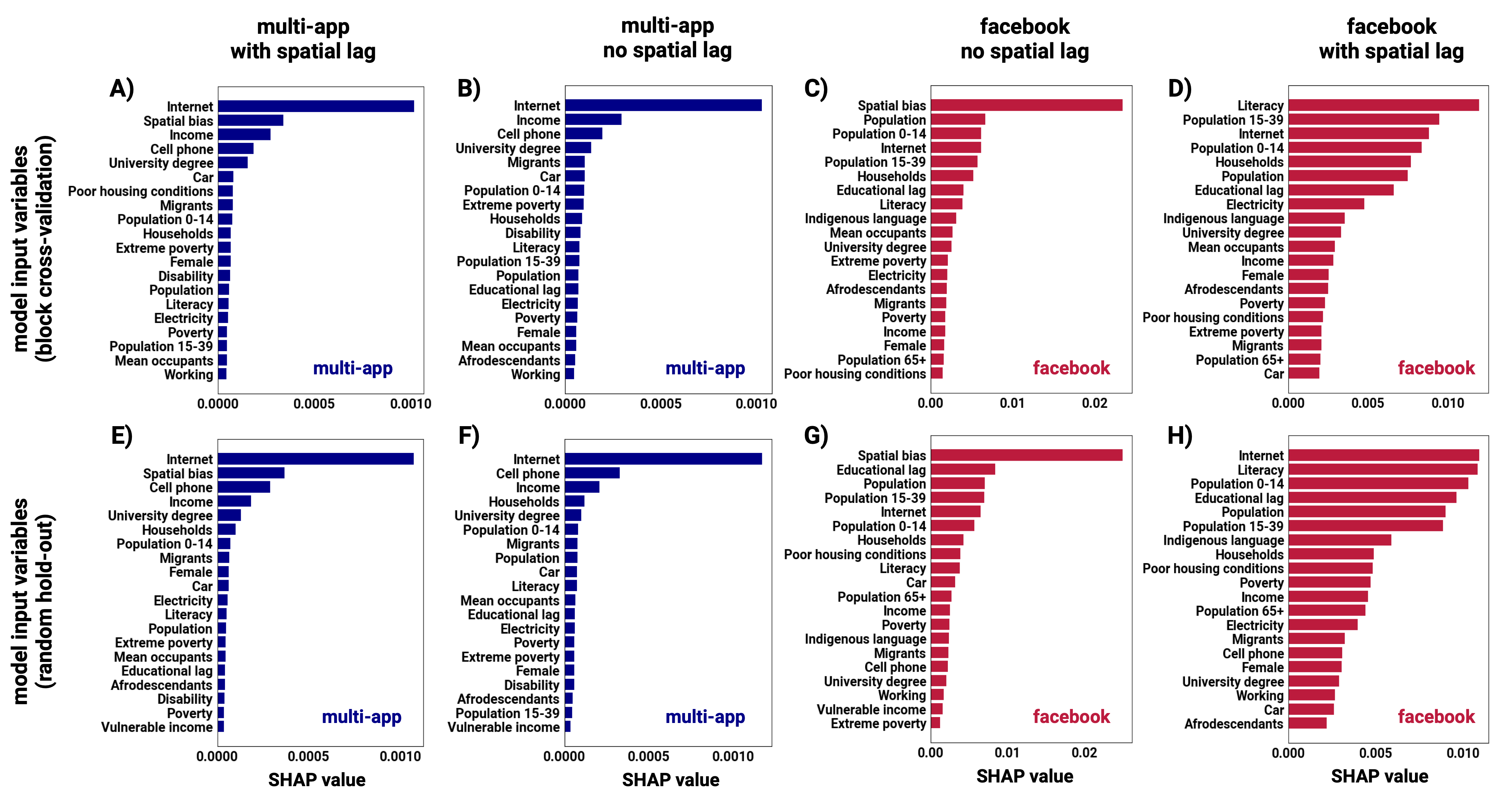}
    \caption{Relative importance of input variables in XGBoost models estimating municipal-level coverage bias for the multi-app (blue) and Facebook (red) datasets. Importance is measured using mean absolute SHAP values. Panels (A–D) show results from models estimated using spatial block cross-validation, while panels (E–H) show results from random hold-out validation with a 70–30 training–test split. For each validation strategy, results are reported for model specifications with and without a spatially lagged bias term. Variables are ordered by decreasing importance within each panel. All models are estimated at the municipal level.}
    \label{fig:importance}
\end{figure}
\end{landscape}

\begin{figure}[h!]
    \centering
    \includegraphics[width=0.7\linewidth]{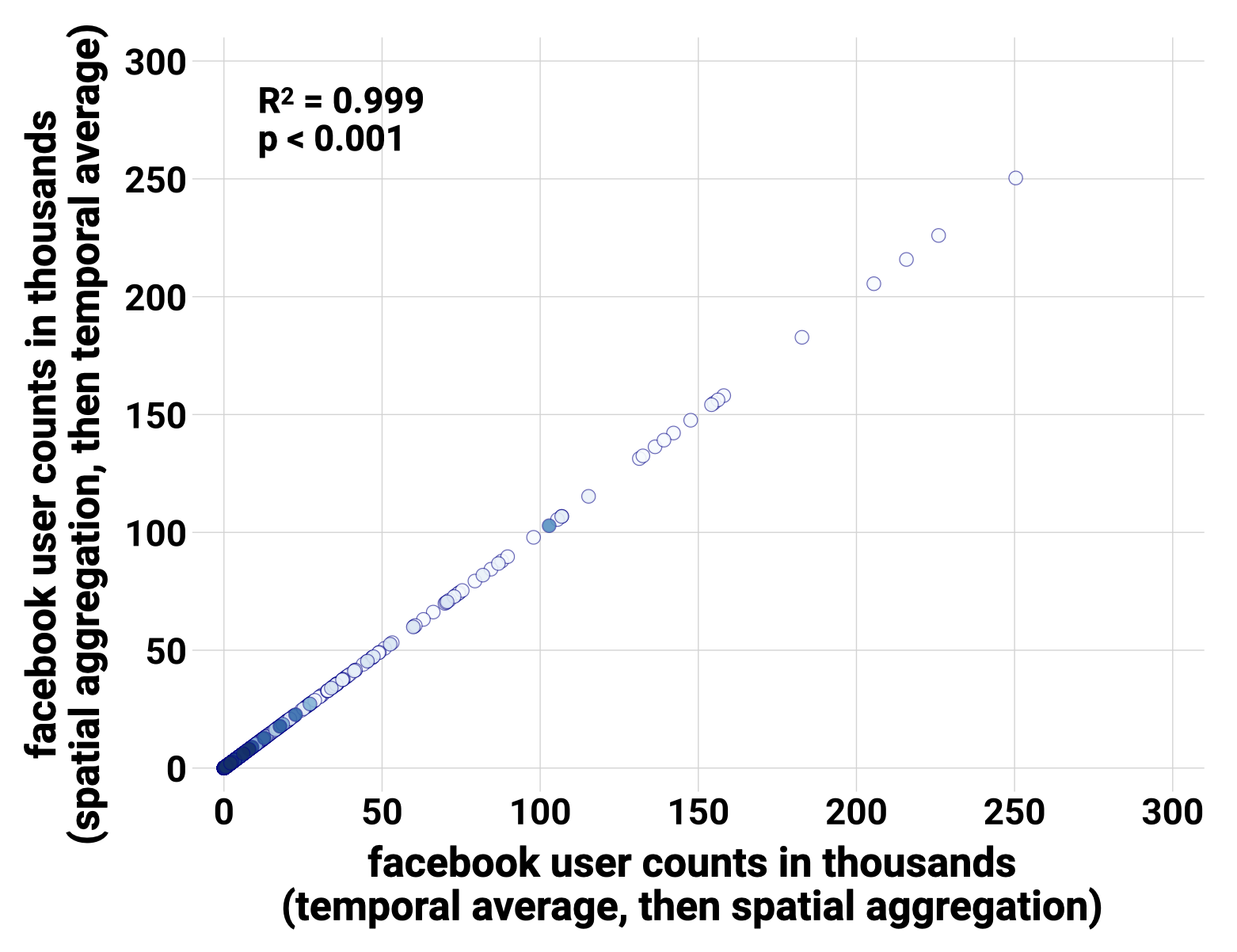}
    \caption{Robustness test of Facebook user counts to the order of temporal and spatial aggregation. The municipal-level Facebook user counts obtained by first computing temporal averages at the tile level and then aggregating spatially to municipalities are shown on the $x$-axis; counts obtained by first aggregating tile-level observations to municipalities for each day and then computing temporal averages are shown on the $y$-axis. Each point represents a municipality and values are reported in thousands of users. The reported $R^2$ and $p$-value between the two aggregation procedures indicates that the order of temporal and spatial aggregation has a negligible effect on estimated population counts.}
    \label{fig:tts-stt}
\end{figure}

\begin{table}[h!]
    \centering
    \includegraphics[width=0.9\linewidth]{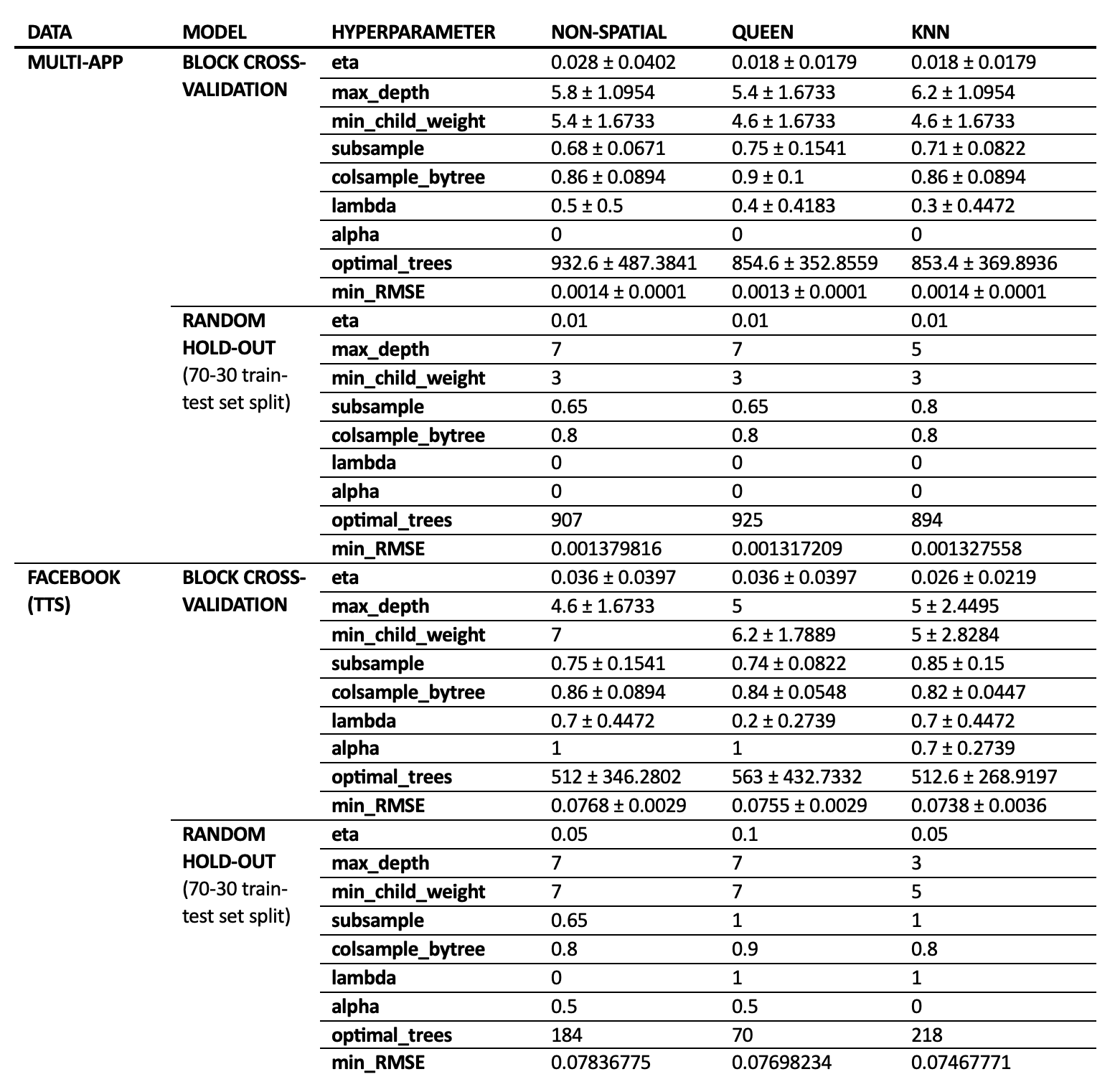}
    \caption{XGBoost model hyperparameters, for multi-app and Facebook data, under spatial block cross-validation and random hold-out validation.}
    \label{tab:hyperparameters}
\end{table}